\documentclass[useAMS]{mn2e}
 \usepackage{times}
 \usepackage{graphicx}
 \usepackage{epsfig}
 \usepackage{amssymb}
 \usepackage{media9}
 \usepackage{hyperref}

 \title[A resonant family of dynamically cold NEOs]
       {A resonant family of dynamically cold small bodies in the near-Earth asteroid belt}

 \author[C. de la Fuente Marcos and R. de la Fuente Marcos]
        {C.~de~la~Fuente~Marcos\thanks{E-mail: nbplanet@fis.ucm.es}
         and
         R. de la Fuente Marcos \\
         Universidad Complutense de Madrid,
         Ciudad Universitaria, E-28040 Madrid, Spain}
 \date{Accepted 2013 May 13.
       Received 2013 May 10;
       in original form 2013 February 25}
 \pubyear{2013}
 
\begin{document}
  \maketitle

  \begin{abstract}
     Near-Earth objects (NEOs) moving in resonant, Earth-like orbits are potentially important. On the positive side, 
     they are the ideal targets for robotic and human low-cost sample return missions and a much cheaper alternative to 
     using the Moon as an astronomical observatory. On the negative side and even if small in size (2--50~m), they have 
     an enhanced probability of colliding with the Earth causing local but still significant property damage and loss of 
     life. Here, we show that the recently discovered asteroid 2013~BS$_{45}$ is an Earth co-orbital, the sixth 
     horseshoe librator to our planet. In contrast with other Earth's co-orbitals, its orbit is strikingly similar to 
     that of the Earth yet at an absolute magnitude of 25.8, an artificial origin seems implausible. The study of the 
     dynamics of 2013~BS$_{45}$ coupled with the analysis of NEO data show that it is one of the largest and most stable 
     members of a previously undiscussed dynamically cold group of small NEOs experiencing repeated trappings in the 1:1 
     commensurability with the Earth. This new resonant family is well constrained in orbital parameter space and it 
     includes at least 10 other transient members: 2003~YN$_{107}$, 2006~JY$_{26}$, 2009~SH$_{2}$ and 2012~FC$_{71}$ 
     among them. 2012~FC$_{71}$ represents the best of both worlds as it is locked in a Kozai resonance and is unlikely 
     to impact the Earth. These objects are not primordial and may have originated within the Venus--Earth--Mars region 
     or in the main-belt, then transition to Amor-class asteroid before entering Earth's co-orbital region. Objects in 
     this group could be responsible for the production of Earth's transient irregular natural satellites.
  \end{abstract}

  \begin{keywords}
     celestial mechanics -- 
     minor planets, asteroids: individual: 2003~YN$_{107}$ --
     minor planets, asteroids: individual: 2006~JY$_{26}$ --
     minor planets, asteroids: individual: 2012~FC$_{71}$ --
     minor planets, asteroids: individual: 2013~BS$_{45}$ --
     planets and satellites: individual: Earth.  
  \end{keywords}

  \section{Introduction}
     During the last two decades, observations of near-Earth Objects (NEOs) have uncovered the existence of a near-Earth asteroid belt made 
     of minor bodies with diameters smaller than 50 m and moving in Earth-like orbits with low eccentricity (Rabinowitz et al. 1993). 
     Putative members of this near-Earth belt have perihelia in the range 0.9--1.1 au, aphelia less than 1.4 au, low eccentricities, a wide 
     range of inclinations and unusual spectral properties (Rabinowitz 1994). This parameter range is sometimes called the Arjuna region
     (Rabinowitz et al. 1993; Gladman, Michel \& Froeschl\'e 2000) after the hero of Hindu epic poem Mahabharata. Main-belt asteroids are 
     not a plausible source for this near-Earth belt (Bottke et al. 1996; Rabinowitz 1997). Amor asteroid fragments provide a reasonable 
     origin for most members and planetary ejecta from Mars, the Earth--Moon system and Venus may have produced the lowest inclination 
     objects (Bottke et al. 1996). Although the vast majority of NEOs have significant eccentricity and/or inclination, a dynamically cold 
     population that includes objects with both low eccentricity and low inclination exists. Periodic close encounters with the Earth--Moon 
     system make this type of orbits quite unstable; therefore, its members are necessarily transient but, due to their Earth-like orbital 
     elements, they may also easily become temporary co-orbitals, in particular horseshoe librators to the Earth. 
     
     The dynamical evolution of asteroids moving in Earth-like orbits has already been studied (Tancredi 1997; Brasser \& Wiegert 2008; 
     Kwiatkowski et al. 2009; Granvik, Vaubaillon \& Jedicke 2012) but objects in resonance with the Earth have been explicitly excluded. 
     However, the existence, characterization and study of such group could be of great importance not only for planning low-cost 
     interplanetary missions (e.g., Davis, Friedlander \& Jones 1993) or asteroid mining (e.g., Lee 2012) but also to reduce the 
     effects of future Earth impact events as those objects are easy to access but, on a dark note, they also have an increased probability 
     of becoming impactors due to their low relative velocities to the Earth (e.g., Lewis 1996). Using data from the JPL HORIZONS 
     system\footnote{http://ssd.jpl.nasa.gov/?horizons} to explore this neglected population, we found that the recently discovered Aten 
     asteroid 2013~BS$_{45}$ exhibits all the orbital attributes expected of an object in such group. Not only its relative (to the Earth) 
     semimajor axis is 0.0024 au but also the values of its eccentricity (0.08) and inclination (0$\fdg$8) are close to those of the 
     Earth. In fact, its orbit is currently the most Earth-like among those of asteroids moving in Earth-like orbits. The study of the 
     dynamical evolution of 2013~BS$_{45}$ led us to find a previously undiscussed dynamically cold group of small near-Earth asteroids
     that experience repeated resonant episodes with the Earth. This Letter is organized as follows: in Section 2, we briefly outline our 
     numerical model. Section 3 focuses on 2013~BS$_{45}$. Section 4 is devoted to the new dynamically cold resonant family. The relevance 
     of our findings is discussed and our conclusions summarized in Section 5. 
%
%
     \begin{table*}
      \fontsize{8}{11pt}\selectfont
      \tabcolsep 0.15truecm
      \caption{Heliocentric Keplerian orbital elements of asteroids 2013 BS$_{45}$, 2003 YN$_{107}$, 2006 JY$_{26}$ and 2012 FC$_{71}$.  
               Values include the 1$\sigma$ uncertainty. The orbit of 2013 BS$_{45}$ is based on 87 observations with a data-arc span 
               of 24 d. The orbits are computed at Epoch JD 245 6400.5 that corresponds to 0:00 \textsc{ut} on 2013 April 18 (J2000.0 ecliptic 
               and equinox. Source: JPL Small-Body Database.)
              }
      \begin{tabular}{cccccc}
       \hline
                                                               &   &   2013 BS$_{45}$            &   2003 YN$_{107}$           
                                                                   &   2006 JY$_{26}$            &   2012 FC$_{71}$              \\
       \hline
        Semimajor axis, $a$ (au)                               & = &   0.997 6309$\pm$0.000 0003 &   0.988 71878$\pm$0.000 00005 
                                                                   &   1.009 863$\pm$0.000 009   &   0.989 53$\pm$0.000 02       \\
        Eccentricity, $e$                                      & = &   0.084 0717$\pm$0.000 0006 &   0.013 9379$\pm$0.000 0003   
                                                                   &   0.083 072$\pm$0.000 011   &   0.087 7$\pm$0.000 2         \\
        Inclination, $i$ ($^{\circ}$)                          & = &   0.786 117$\pm$0.000 005   &   4.321 08$\pm$0.000 03       
                                                                   &   1.439 32$\pm$0.000 07     &   4.967$\pm$0.014             \\
        Longitude of the ascending node, $\Omega$ ($^{\circ}$) & = &  85.388 2$\pm$0.000 6       & 264.431 61$\pm$0.000 08       
                                                                   &  43.487$\pm$0.004           &  38.708$\pm$0.006             \\
        Argument of perihelion, $\omega$ ($^{\circ}$)          & = & 146.064 1$\pm$0.000 7       &  87.516 70$\pm$0.000 13       
                                                                   & 273.571$\pm$0.012           & 347.76$\pm$0.02               \\
        Mean anomaly, $M$ ($^{\circ}$)                         & = & 346.006 0$\pm$0.000 2       & 254.342 2$\pm$0.000 5         
                                                                   & 223.70$\pm$0.05             & 186.20$\pm$0.04               \\
        Perihelion, $q$ (au)                                   & = &   0.913 7584$\pm$0.000 0007 &   0.974 9381$\pm$0.000 0002   
                                                                   &   0.925 972$\pm$0.000 003   &   0.902 8$\pm$0.000 2         \\
        Aphelion, $Q$ (au)                                     & = &   1.081 5034$\pm$0.000 0003 &   1.002 49948$\pm$0.000 00005 
                                                                   &   1.093 755$\pm$0.000 009   &   1.076 29$\pm$0.000 02\\
        Absolute magnitude, $H$ (mag)                          & = &  25.8$\pm$0.3             &  26.3$\pm$0.7               
                                                                   &  28.3$\pm$0.6             &  25.2$\pm$0.4               \\
       \hline
      \end{tabular}
      \label{elements}
     \end{table*}
%
%
  \section{Numerical model}
     The numerical integrations of the orbits of the objects studied here were performed with the Hermite integrator (Makino 1991; Aarseth 
     2003), in a model Solar system which takes into account the perturbations by eight major planets and treats the Earth--Moon system as 
     two separate objects; it also includes the barycentre of the dwarf planet Pluto--Charon system and the three largest asteroids (for 
     further details, see de la Fuente Marcos \& de la Fuente Marcos 2012). Results in the figures have been obtained using initial 
     conditions (positions and velocities in the barycentre of the Solar system) provided by the JPL HORIZONS system (Giorgini et 
     al. 1996; Standish 1998) and referred to the JD 245 6400.5 epoch which is the $t$ = 0 instant. In addition to the calculations 
     completed using the nominal orbital elements in Table \ref{elements}, we have performed 50 control simulations for each object with 
     sets of orbital elements obtained from the nominal ones within the accepted uncertainties (3$\sigma$).
%
%
     \begin{figure}
        \centering
        \includemedia[
          label=2013BS45,
          width=\linewidth,height=0.65\linewidth,
          activate=onclick,
          addresource=2013BS45w.mp4,
          flashvars={
            source=2013BS45w.mp4
           &autoPlay=true
           &loop=true
           &controlBarMode=floating
           &controlBarAutoHide=false
           &scaleMode=letterbox
          }
        ]{\includegraphics{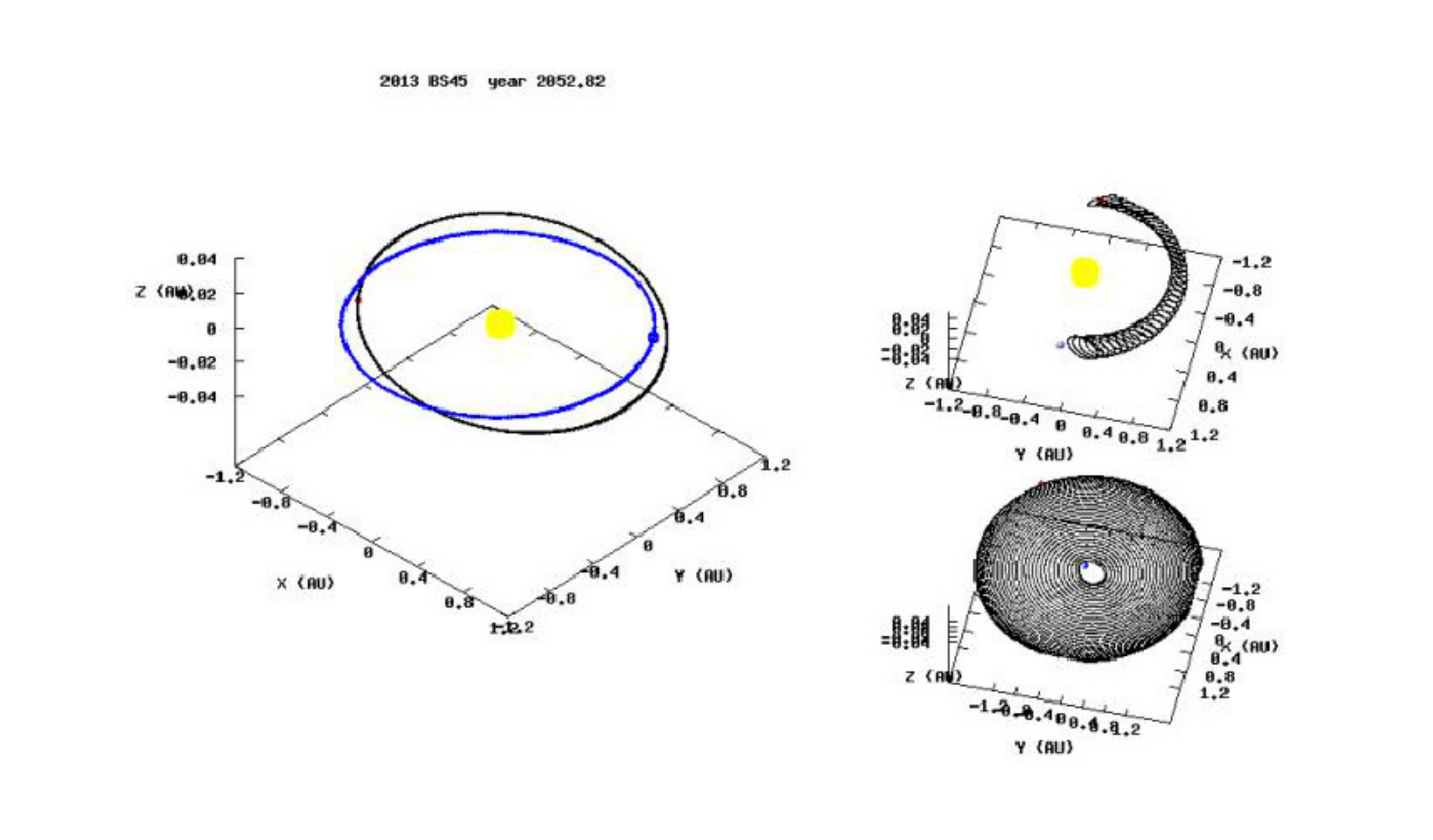}}{VPlayer.swf}
        \PushButton[
           onclick={
             annotRM['2013BS45'].activated=true;
             annotRM['2013BS45'].callAS('play');
           }
        ]{\fbox{Play}}
        \PushButton[
           onclick={
             annotRM['2013BS45'].activated=true;
             annotRM['2013BS45'].callAS('pause');
           }
        ]{\fbox{Pause}}
        \caption{Three-dimensional evolution of the orbit of 2013~BS$_{45}$ in three different frames of reference: heliocentric (left),
                 frame corotating with the Earth but centred on the Sun (top right) and geocentric (bottom right). The red point marks
                 2013~BS$_{45}$, the blue one the Earth and the yellow one the Sun. The osculating orbits are outlined and the viewing 
                 angle changes slowly to facilitate visualizing the orbital evolution.}
        \label{animation1}
     \end{figure}
%
%

  \section{2013~BS$_{45}$, signalling the dynamically cold population}
\begin{figure*}
       \centering
        \includegraphics[width=\linewidth]{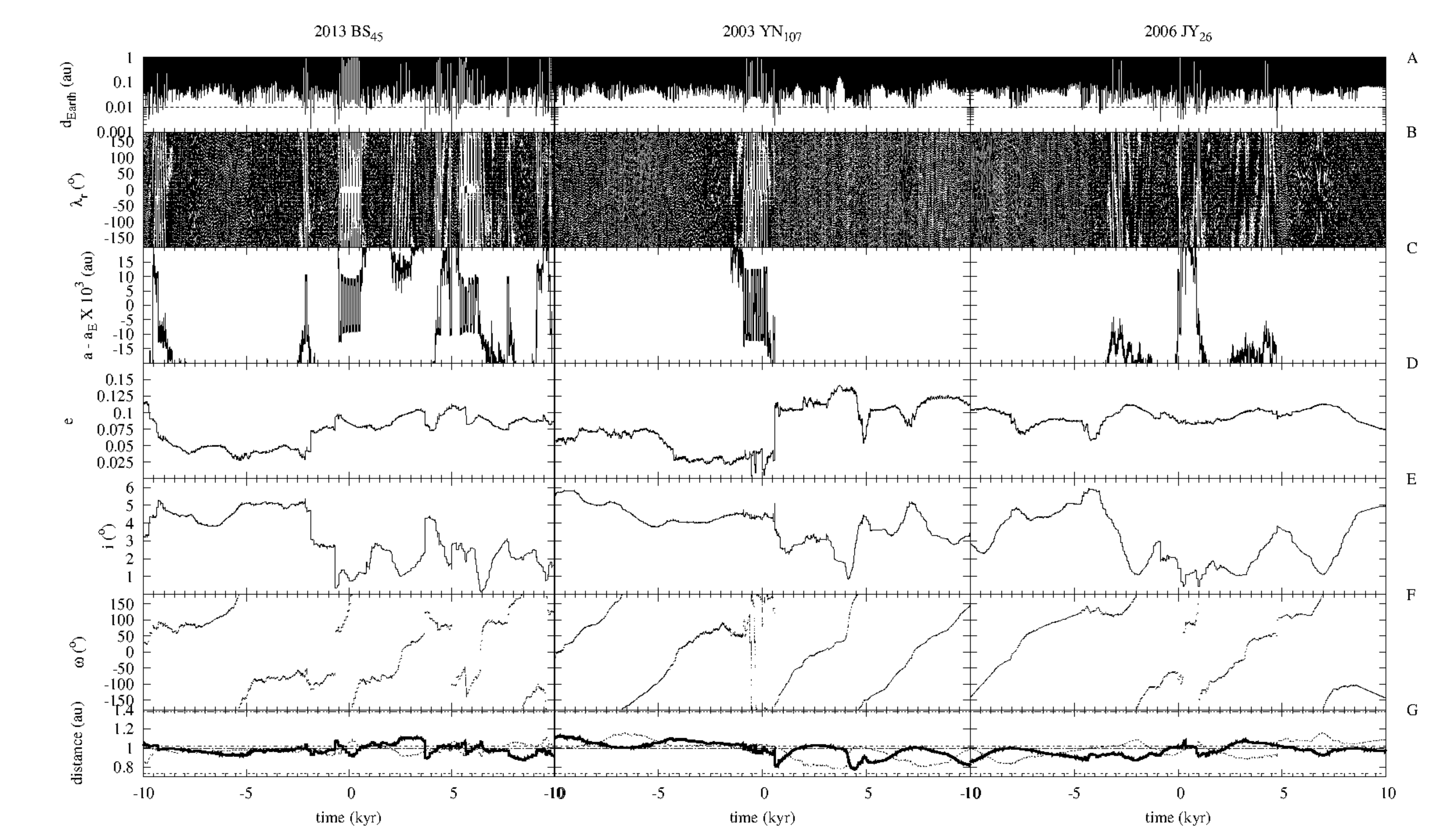}
        \caption{Comparative time evolution of various parameters for active, near-Earth, dynamically cold horseshoe librators 2013~BS$_{45}$ 
                 (left), 2003 YN$_{107}$ (centre) and 2006 JY$_{26}$ (right). The distance from the Earth (panel A); the value of the Hill 
                 sphere radius of the Earth, 0.0098 au, is displayed. The resonant angle, $\lambda_{\rm r}$ (panel B) for the orbit in Table 
                 \ref{elements}. The orbital elements $a - a_{\rm Earth}$ (panel C), $e$ (panel D), $i$ (panel E) and $\omega$ (panel F). 
                 The distance to the descending (thick line) and ascending nodes (dotted line) is in panel G. Mars' and Earth's perihelion 
                 and Earth's and Venus' aphelion distances are also shown.
                }
        \label{all}
     \end{figure*}

     2013~BS$_{45}$ was discovered on 2013 January 20 by J.~V. Scotti observing with the Steward Observatory 0.9-m Spacewatch telescope at 
     Kitt Peak (Bressi et al. 2013) and had a close encounter with the Earth on 2013 February 12 at 0.013 au. It is small with $H$ = 25.8 
     which translates into a diameter in the range 20--40 m for an assumed albedo of 0.20--0.04. Radar observations indicate that it may be 
     a very rapid rotator with a period of just a few minutes\footnote{http://echo.jpl.nasa.gov/asteroids/2013BS45/2013BS45\_planning.html} 
     pointing to relatively recent collisional debris. Even if small, the object is much larger than the previously known tiny (so-called) 
     minimoons 1991~VG (Tancredi 1997) and 2006~RH$_{120}$ (Kwiatkowski et al. 2009). The orbital elements of 2013~BS$_{45}$ (see Table
     \ref{elements}) are suggestive of a NEO co-orbital with the Earth, likely of the horseshoe kind. In order to confirm its co-orbital 
     nature, we have performed $N$-body calculations in both directions of time. The three-dimensional evolution of its orbit for several 
     decades is shown in Fig. \ref{animation1}. Such a path viewed in a frame of reference rotating with the Earth looks like a corkscrew 
     around the orbit of the host planet while both revolve around the Sun. Regular horseshoe orbiters are characterized by the libration of 
     the difference between the mean longitudes of the object and its host planet or relative mean longitude, $\lambda_{\rm r}$. The mean 
     longitude of an object is given by $\lambda$ = $M$ + $\Omega$ + $\omega$, where $M$ is the mean anomaly, $\Omega$ is the longitude of 
     ascending node and $\omega$ is the argument of perihelion. If the libration amplitude is larger than 180$^{\circ}$, encompassing 
     L$_3$, L$_4$ and L$_5$ but not reaching the actual planet, it is said that the object follows a symmetric horseshoe orbit (Murray \& 
     Dermott 1999). The first minor body confirmed to follow a horseshoe orbit (with the Earth or any other planet) was 3753~Cruithne 
     (1986~TO) (Wiegert, Innanen \& Mikkola 1997, 1998). Connors et al. (2004) identified a second horseshoe librator with the Earth in 
     2003~YN$_{107}$. Additional objects moving in comparable trajectories are 2002~AA$_{29}$, 2001~GO$_{2}$ (Brasser et al. 2004) and 
     2010~SO$_{16}$ (Christou \& Asher 2011). Therefore, 2013~BS$_{45}$ is the sixth horseshoe librator to our planet following a path 
     qualitatively similar to that of the first known asteroidal companion to the Earth, 3753. The details are different because, in sharp 
     contrast with 2013~BS$_{45}$, the orbit of 3753 is quite eccentric ($e$ = 0.5) and inclined ($i$ = 19$\fdg$8). The current orbit 
     of 2013~BS$_{45}$ (see Table \ref{elements}) is very reliable; based on 87 observations (including 4 radar observations, 1 Doppler and 
     3 delays from Goldstone$^2$) with a data-arc span of 24 d, we can state with certainty that it is a horseshoe orbiter. The longer term 
     evolution of its orbital parameters appears in Fig. \ref{all}; it will remain moving in a symmetric horseshoe orbit for about 1000 yr 
     with a period of nearly 160 yr. Control orbits give consistent results and they reveal that 2013~BS$_{45}$ has experienced co-orbital 
     episodes in the past and it will return to the 1:1 commensurability in the future. When trapped in the 1:1 resonance with the Earth, 
     its orbital elements evolve in a restricted area of parameter space. The implications of this finding are studied next. 
%
%
%
%

  \section{A resonant family of dynamically cold small bodies}
\begin{figure}
       \centering
        \includegraphics[width=1.05\linewidth]{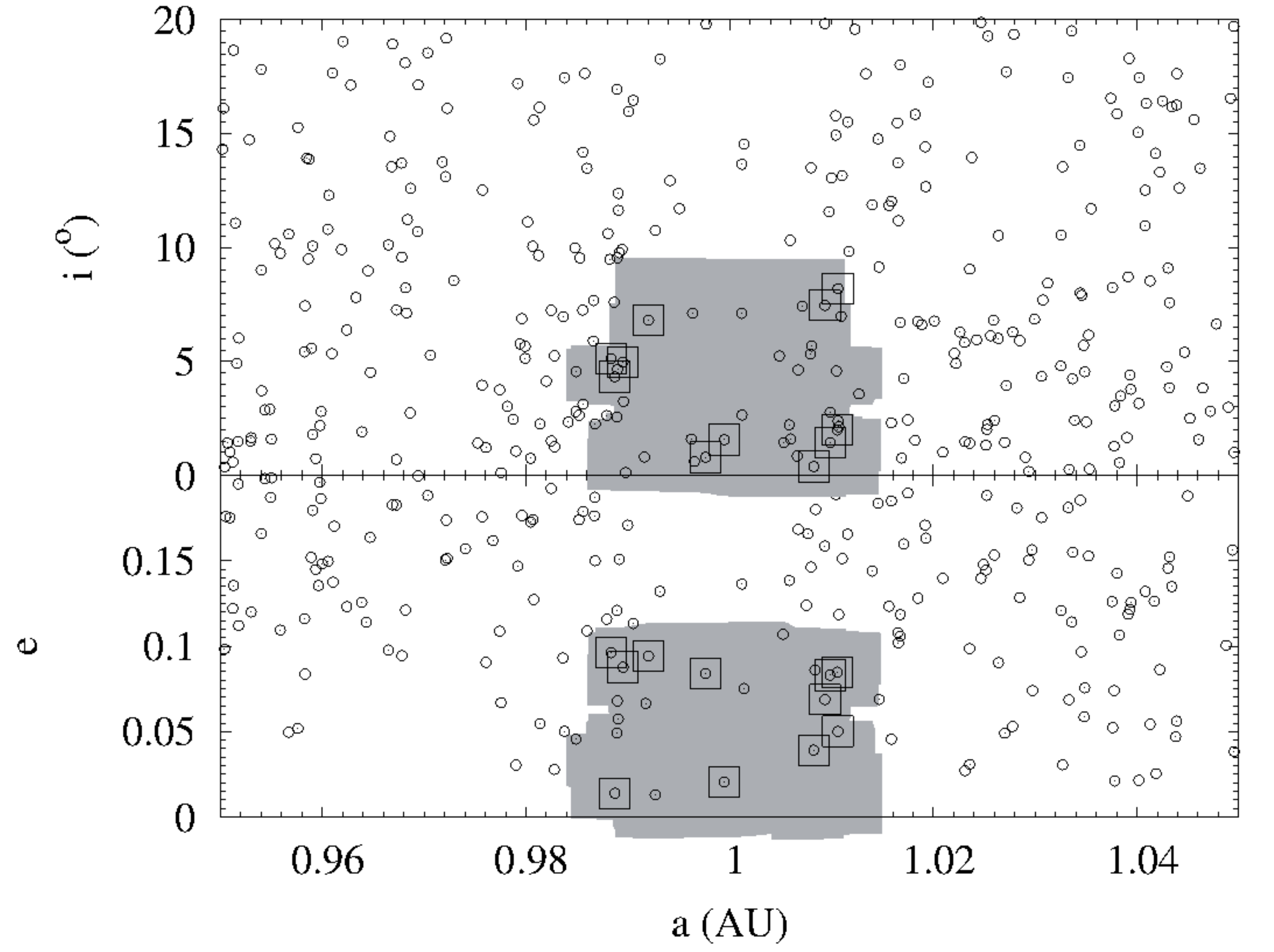}
        \caption{Orbital parameter space (grey area) occupied by the dynamically cold resonant family studied here.  
                 Circles are data from the JPL HORIZONS system; squares show the objects in Table 2.
                }
        \label{region}
     \end{figure}

\begin{figure}
       \centering
        \includegraphics[width=1.05\linewidth]{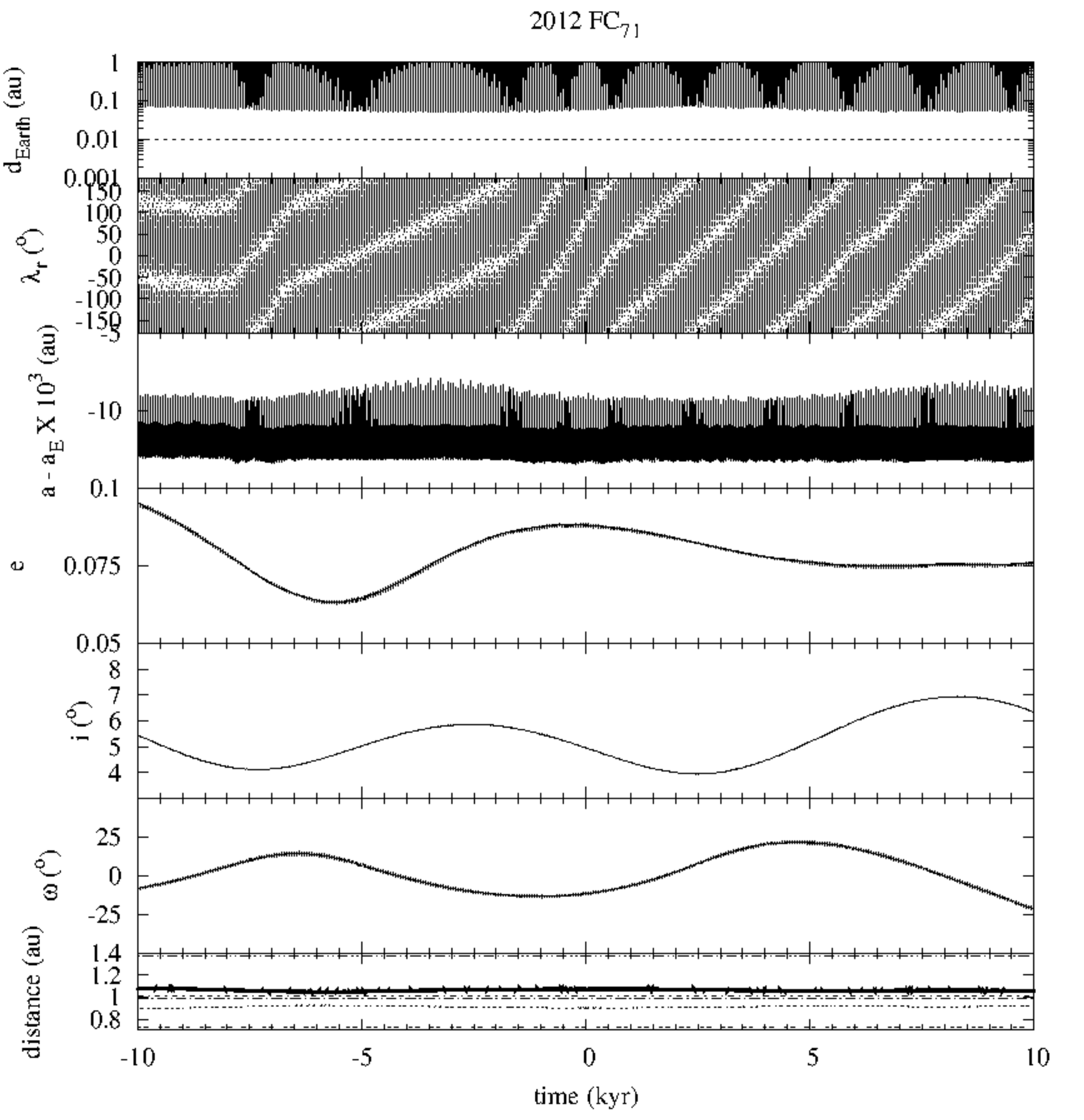}
        \caption{Similar to Fig. \ref{all} but for 2012 FC$_{71}$, a Kozai resonator.
                }
        \label{2012FC71}
     \end{figure}

\begin{figure}
        \centering
        \includemedia[
          label=2012FC71,
          width=\linewidth,height=0.65\linewidth,
          activate=onclick,
          addresource=2012FC71w.mp4,
          flashvars={
            source=2012FC71w.mp4
           &autoPlay=true
           &loop=true
           &controlBarMode=floating
           &controlBarAutoHide=false
           &scaleMode=letterbox
          }
        ]{\includegraphics{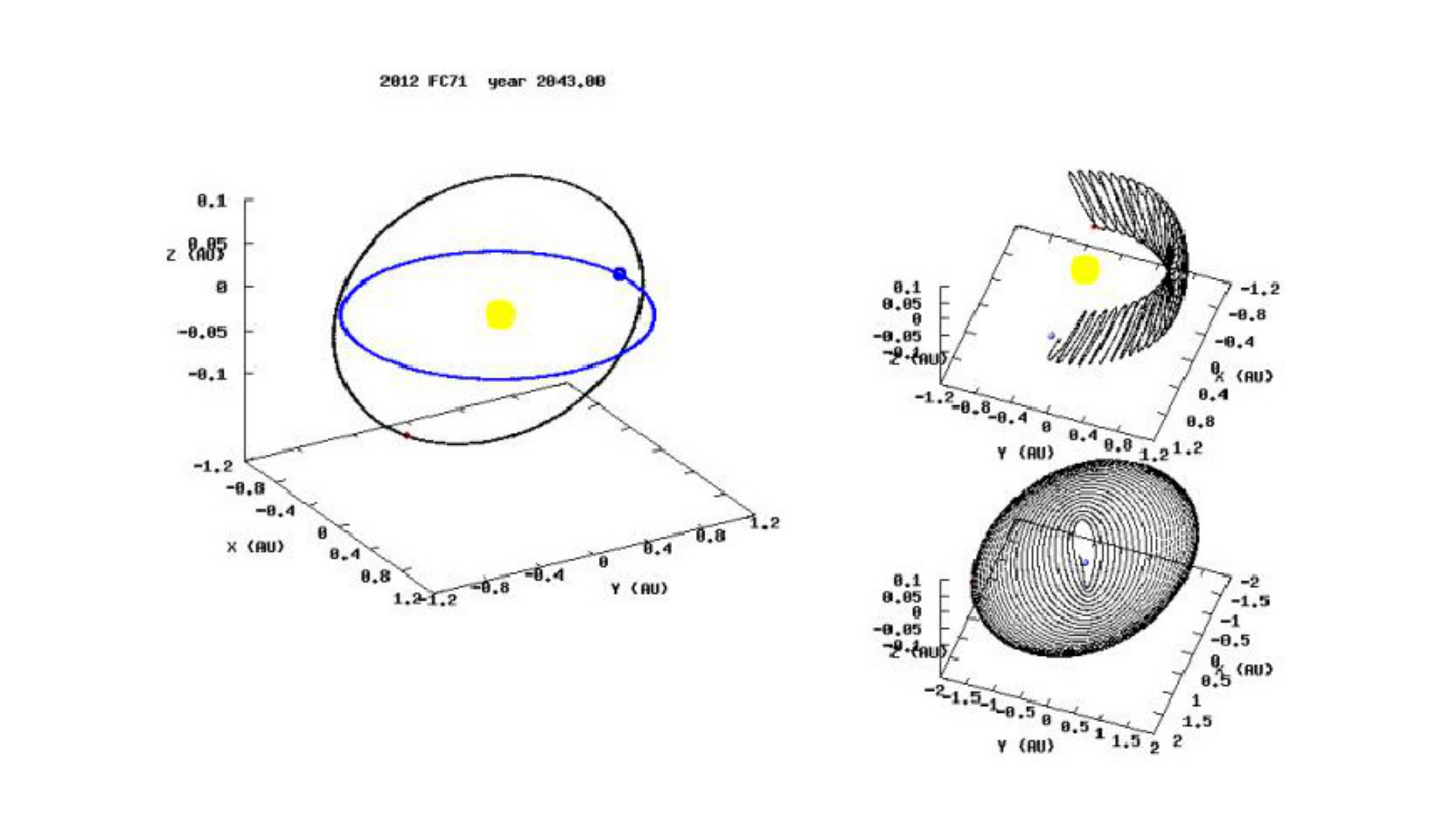}}{VPlayer.swf}
        \PushButton[
           onclick={
             annotRM['2012FC71'].activated=true;
             annotRM['2012FC71'].callAS('playPause');
           }
        ]{\fbox{Play/Pause}}
        \caption{Similar to Fig. \ref{animation1} but for 2012 FC$_{71}$.}
        \label{animation2}
     \end{figure}

     The study of the dynamics of 2013~BS$_{45}$ clearly exposed that, during the co-orbital episodes, its path was constrained to a well 
     defined region in the orbital parameter space, in particular $a$, $e$ and $i$: 0.985 $< a$ (au) $<$ 1.013, 0 $< e <$ 0.1 and 0 $< i 
     <$ 8$\fdg$56. During these episodes, the values of the osculating orbital elements $a$, $e$ and $i$ remain restricted to the grey 
     areas in Fig. \ref{region}. The boundary is not regular likely because of secular resonances. The results from multiple simulations 
     have been used to outline the region of interest. The level of stability is not the same within the entire region. This behaviour 
     motivated a search for additional objects within that well-defined volume in $a$--$e$--$i$ space that produced 10 candidates for 
     membership in our suspected dynamical family (see Table \ref{members}). After performing additional $N$-body calculations analogous to 
     those completed for 2013~BS$_{45}$, we found that 2009~SH$_{2}$, 2003~YN$_{107}$ and 2006~JY$_{26}$ currently are horseshoe librators 
     to the Earth. The orbital evolution of 2003~YN$_{107}$ and 2006~JY$_{26}$ is shown in Fig. \ref{all}. As pointed out above, 
     2003~YN$_{107}$ was identified as an Earth co-orbital shortly after its discovery (Brasser et al. 2004; Connors et al. 2004). Our 
     calculations indicate that it is mainly a regular horseshoe librator not a quasi-satellite but it is currently ending a very brief 
     quasi-satellite episode ($\sim$15~yr). The remaining objects underwent brief horseshoe episodes in the past or will do in the near 
     future. 2009~SH$_{2}$ started a horseshoe episode about 30 yr ago and it will leave its co-orbital state in about 100 yr from now. 
     2010~HW$_{20}$ is not currently co-orbital but it will become one, similar to 2003~YN$_{107}$, in a few hundred years. A similar 
     behaviour is observed for 2012~LA$_{11}$. 2008~KT will become co-orbital in about 2000 yr. 2008~UC$_{202}$ will be co-orbital in 600 yr. 
     2009~BD was a horseshoe librator about 400 yr ago and it will repeat in about 1400 yr. 2006~JY$_{26}$ is currently a horseshoe librator, 
     leaving the state in about 130 yr. All these objects are capable of co-orbital evolution yet their orbits are very different from those 
     of 3753 (Wiegert et al. 1997) or 2010~SO$_{16}$ (Christou \& Asher 2011), well-studied Earth co-orbitals. The objects that are 
     currently horseshoe librators are the transient component of the resonant group but we found a surprising meta-stable component too. 
     Asteroid 2012~FC$_{71}$ is locked in a Kozai resonance (Kozai 1962) with the argument of perihelion librating around 0$^{\circ}$ (see 
     Fig. \ref{2012FC71}). This object is, at a proper average inclination of 5$\fdg$5, the coldest known Kozai resonator. In sharp 
     contrast with horseshoe librators, NEOs trapped in the Kozai resonance have a very slow orbital evolution (see Fig. \ref{2012FC71}) and 
     can remain relatively unperturbed for hundreds of thousands of years (Michel \& Thomas 1996) as they never get closer than 0.07 au to 
     the Earth. They do not librate and are not real co-orbitals although their paths resemble those of true co-orbitals (compare Figs. 
     \ref{animation1} and \ref{animation2}). 2006~RH$_{120}$ can also experience co-orbital motion but more frequently and in the future, 
     Kozai episodes. The objects in this dynamical family exhibit relatively short, recurrent co-orbital episodes with the Earth or are
     low-inclination--low-eccentricity Kozai resonators, easily switching between the various resonant states as a result of close 
     encounters with the Earth--Moon system. Like the Hilda family (Bro\v{z} \& Vokrouhlick\'y 2008), this is a dynamical group or resonant 
     family not a genetic one as the objects are unlikely to descend from a common parent body. 
%
%
     \begin{table*}
      \fontsize{8}{10pt}\selectfont
      \tabcolsep 0.35truecm
      \caption{Orbital properties of the dynamically cold members of the near-Earth asteroid belt  
               (source for $\Delta v$:  http://echo.jpl.nasa.gov/$\sim$lance/delta\_v/delta\_v.rendezvous.html).
              }
      \begin{tabular}{ccccccccc}
       \hline
         Designation     & $a$ (au)  &   $e$     & $i$ ($^{\circ}$) & $\Omega$ ($^{\circ}$) & $\omega$ ($^{\circ}$) & $H$ (mag) & $U$       & $\Delta v$ (km s$^{-1}$) \\
       \hline
         2009 SH$_{2}$   & 0.992 035 & 0.094 195 &   6.810 19       &     101.483 37        &        6.720 39       &   24.90   & 0.150 998 &  5.070 \\
         2012 FC$_{71}$  & 0.989 531 & 0.087 678 &   4.967 37       &     347.757 12        &       38.707 67       &   25.23   & 0.122 561 &  4.696 \\
         2013 BS$_{45}$  & 0.997 631 & 0.084 072 &   0.786 12       &     146.064 22        &       85.388 07       &   25.85   & 0.085 177 &  4.045 \\
         2010 HW$_{20}$  & 1.010 623 & 0.050 118 &   8.188 72       &      60.114 82        &       39.251 30       &   26.10   & 0.151 383 &  5.690 \\
         2012 LA$_{11}$  & 0.988 356 & 0.096 080 &   5.115 55       &     242.277 07        &      261.209 38       &   26.15   & 0.130 298 &  4.751 \\
         2003 YN$_{107}$ & 0.988 719 & 0.013 938 &   4.321 08       &      87.516 70        &      264.431 61       &   26.28   & 0.075 816 &  4.879 \\
         2008 KT         & 1.010 578 & 0.084 823 &   1.984 36       &     102.037 34        &      240.642 59       &   28.21   & 0.091 460 &  4.425 \\
         2008 UC$_{202}$ & 1.009 357 & 0.068 715 &   7.458 96       &      91.258 36        &       37.403 81       &   28.24   & 0.147 130 &  5.479 \\
         2009 BD         & 1.008 256 & 0.039 180 &   0.381 33       &     113.468 17        &       59.394 10       &   28.24   & 0.039 195 &  3.870 \\
         2006 JY$_{26}$  & 1.009 863 & 0.083 072 &   1.439 32       &     273.570 56        &       43.487 20       &   28.35   & 0.086 644 &  4.364 \\
         2006 RH$_{120}$ & 0.999 473 & 0.020 577 &   1.561 01       &     183.400 42        &      290.597 99       &   29.53   & 0.034 132 &  3.813 \\
       \hline
      \end{tabular}
      \label{members}
     \end{table*}
%
%
%
%
%
%
%
%
%
%
%
%
%
%

  \section{Discussion and conclusions}
     It should be noticed that the orbits of these objects are highly chaotic, with e-folding times of 10--100 yr, but they remain as NEOs 
     for nearly 1 Myr. Only a fraction of that time, typically $<$ 10\,000 yr, the objects stay in the region defined above but in 
     some cases they remain there for nearly 0.3 Myr. Although the current dynamical status of these objects is reliable, predictions beyond 
     a few hundred years can only be made in statistical terms due to frequent close encounters, well below the Hill radius (0.0098 au), 
     with the Earth--Moon system (see Fig. \ref{all}, panel A). Objects in this group may experience repeated co-orbital and Kozai episodes 
     and transitions are triggered by the encounters. The most usual state is symmetric horseshoe but quasi-satellite and Trojan episodes 
     have also been observed during the simulations. Due to their low relative velocities to the Earth most of the objects studied here 
     undergo a large number of brief episodes (1--30 d) in which their Keplerian geocentric energy becomes negative even if they are 
     several Hill radii from the Earth. In Table \ref{members}, the value $U$ is the dimensionless encounter velocity with respect to the 
     Earth at infinity ($U = v_{\rm r}/v_{\rm E}$, where $v_{\rm r}$ is the relative velocity between the object and the Earth, and 
     $v_{\rm E}$ is the circular velocity of the Earth at 1 au), given by $U = \sqrt{3 - T_{\rm E}}$, where $T_{\rm E}$ is the Tisserand 
     parameter related to the Earth. All the objects have very low values of $U$. 2006~RH$_{120}$ was an actual natural satellite of the 
     Earth for over a year in 2006 (Kwiatkowski et al. 2009): it has experienced similar episodes in the past and it will probably repeat 
     them in the future. Besides 2006~RH$_{120}$, the only other object that has been involved in temporary capture events as defined by 
     Granvik et al. (2012) is 2009~BD but only during a few days; both objects have the lowest values of $U$. The existence of this 
     dynamical group provides a natural source for the so-called Earth's irregular natural satellites. Our analysis suggests that many other 
     objects currently in the neighbourhood of the volume of the orbital parameter space described above (see Fig. \ref{region}, grey areas) 
     may be former members of this dynamical group, like 1991~VG, or may join it in the future. It is not surprising that this dynamically 
     cold family has not been reported earlier. Hidden in plain sight, all the proposed members in Table \ref{members} have been discovered 
     during the last decade. They are small or very small objects with typical maximum apparent magnitude not less than 18 and can only be 
     observed when they get close to the Earth, every 50--80 yr (twice per horseshoe period), remaining within 0.5 au from the Earth for 
     about 5--10 yr with typical minimum distances of 0.01 au. Only during these favourable windows of opportunity are these objects 
     relatively easy to observe, easily accessible, and have risk of impact. In terms of observability, the situation is even worst for the 
     Kozai resonators as they never get that close to the Earth (see Fig. \ref{2012FC71}) and encounters typically occur every 55 yr; if all 
     of them are like 2012~FC$_{71}$ or smaller, then the vast majority are well beyond reach of current NEO surveys and they may be very 
     numerous. 

     Although the objects described here have sizes in the range 2--50 m, they are still large enough to provoke a significant amount of 
     local destruction if they enter the atmosphere. On the positive side, these objects are also the ideal targets (see $\Delta v$ in Table 
     \ref{members}, $\Delta v$ is the minimum total variation of speed for transferring from low-Earth orbit to rendezvous with the object, 
     the equivalent $\Delta v$ to reach the Moon is 6 km s$^{-1}$) for commercial mining of minerals in outer space (e.g., Lee 2012). 
     They can also be used to reach the far side of Earth's orbit, the Lagrangian point L$_3$, at basically no cost. This has multiple 
     astronomical applications for solar-powered, fully automated observatories. These observations can enable the study of the NEO 
     population from the inside, finding about possible impactors well before they get close enough to become a threat. In their way to 
     L$_3$, these objects also visit L$_4$ and L$_5$ where primordial material in the form of dynamically cold, very small Trojans may still 
     be trapped. Instead of using the Moon as an extraterrestrial astronomical observatory, they represent a much cheaper and flexible 
     alternative. Minor bodies moving in orbits with low-eccentricity, low-inclination and Earth-like period are sometimes called Arjuna 
     asteroids (Lewis 1996; Lee 2012). Further observations are needed to expand the list of objects moving in Arjuna-type orbits. 

  \section*{Acknowledgements}
     The authors would like to thank S. J. Aarseth for providing the code used in this research and the referee for his/her constructive and 
     useful report. This work was partially supported by the Spanish `Comunidad de Madrid' under grant CAM S2009/ESP-1496. We thank M. J. 
     Fern\'andez-Figueroa, M. Rego Fern\'andez and the Department of Astrophysics of the Universidad Complutense de Madrid (UCM) for 
     providing computing facilities. Most of the calculations and part of the data analysis were completed on the `Servidor Central de 
     C\'alculo' of the UCM and we thank S. Cano Als\'ua for his help during this stage. In preparation of this Letter, we made use of the 
     NASA Astrophysics Data System, the ASTRO-PH e-print server and the MPC data server.

\end{document}